\documentclass[
    ,final            
  ]
  {aipproc}

\layoutstyle{6x9}

\begin{document}

\title{Prompt photon hadroproduction \\in the $k_T-$factorization
approach}


\author{S.P. Baranov}{
  address={P.N. Lebedev Physics Institute, 
119991 Moscow, Russia}
}

\author{A.V. Lipatov}{
 address={SINP, Moscow State University, 119991 Moscow, Russia}
}

\author{N.P. Zotov}{
address={SINP, Moscow State University, 119991 Moscow, Russia}
}

\begin{abstract}
We study the production of prompt photons at high energy in the framework
of the $k_T-$factorization approach. The amplitude for production of a
single photon associated with
quark pair in the fusion of two off-shell gluons is calculated.
 Theoretical results are compared with the Tevatron data.

\end{abstract}

\maketitle


\section{Introduction}
\indent

The production of prompt photons in hadron-hadron collisions at the 
Tevatron is a subject of intense studies (see, for example, \cite{UB}).
 At the leading order of QCD, prompt photons
can be produced via quark-gluon Compton scattering or quark-antiquark
annihilation and so, the cross sections of these processes are
strongly sensitive to the parton (quark and gluon) content of a proton.
Besides that the events with an isolated photon are an important tool
to study hard interaction processes since such photons
emerge without the hadronization phase.
In standard QCD, the disagreement between experimental data at 
the Tevatron~\cite{d0,cdf,d06} and 
theoretical description (see Refs. in~\cite{UB}) is attributed usually to  
the  initial-state soft-gluon radiation or to the 
intrinsic nonperturbative transverse momentum $k_T$
of the incoming partons.

In the framework of the $k_T$-factorization approach~\cite{sha} the 
treatment of $k_T$-enhancement and gluon emission is more reasonable. In 
this approach the transverse momentum of incoming partons
is generated in the course of non-collinear parton evolution
 of the BFKL type~\cite{BFKL}. In  paper~\cite{LZ}
to analyse the data~\cite{d0,cdf,d06}  the proper 
off-shell expressions for partonic matrix elements 
and  KMR  unintegrated parton densities~\cite{KMRW} have
been used. An important component of the calculations~\cite{LZ} is
 using the unintegrated quark distributions in a proton. At present
these densities are available in the framework of KMR approach only.
It makes difficulties for the invistigation of dependence of the 
calculated cross sections on the non-collinear evolution scheme in the  
$k_T$-factorization approach.

\begin{figure}[!b]
  \includegraphics[height=.27\textheight]{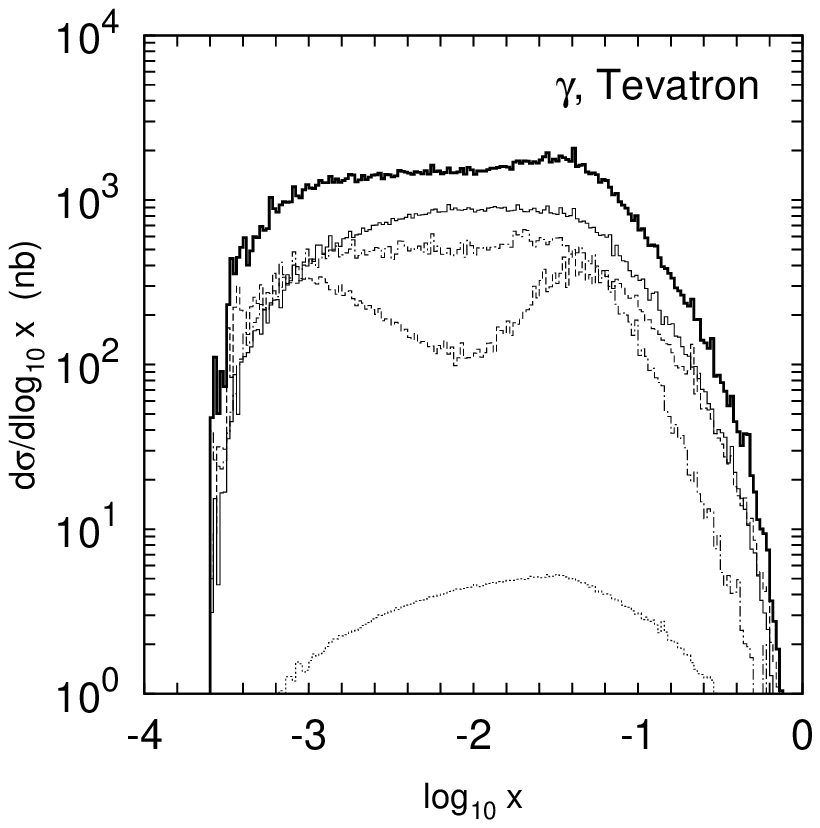}
  \includegraphics[height=.27\textheight]{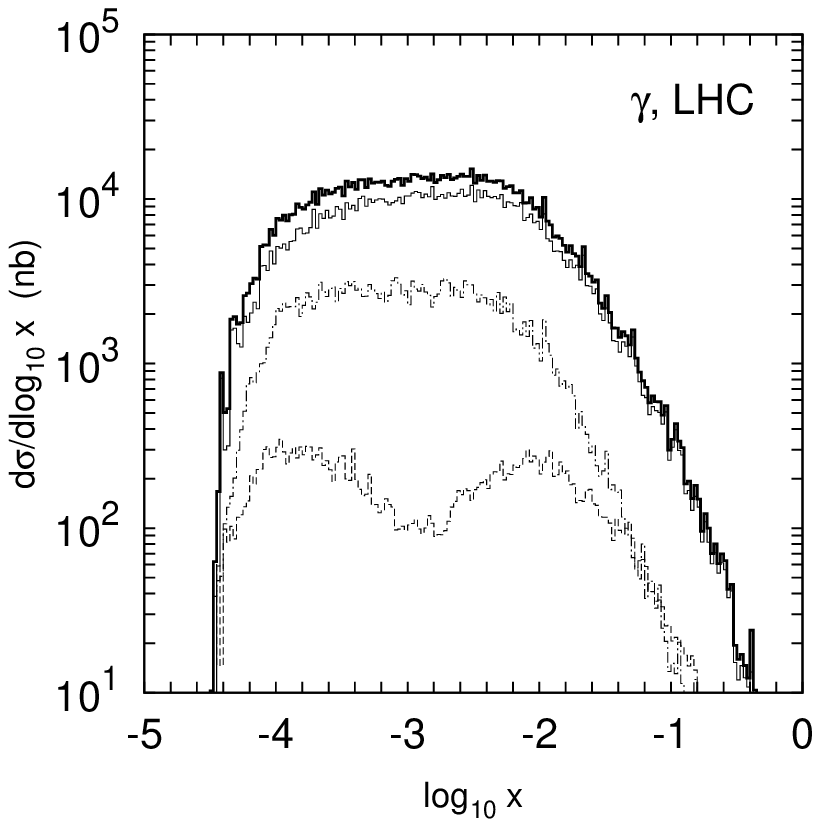}
  \caption{Differential cross section of prompt photon production 
at the Tevatron and LHC as a function of $\log _{10}x$. Different
histograms correspond to different subprocess (see text).
  }
  \label{fig1}
\end{figure}

  In our paper~\cite{BLZ} we have used a different way. Instead of using  
the unintegrated quark distributions and the
corresponding quark-gluon fusion and/or quark-antiquark annihilation cross
sections we have calculated off-shell matrix element of
the $g^* g^* \to q \bar q \gamma$ subprocess, having
 hope for operating in terms of the unintegrated gluon
densities only. But, first of all, these matrix elements cover only
the sea quark contribution.
 However, the contribution from the valence quarks is significant only at 
large $x$, and therefore can be safely taken into account in the collinear
 LO approximation as an additional one. Secondly, this idea can 
only work well if the sea quarks appear from the last step of the gluon 
evolution. 
 This method does not apply to the quarks 
coming from the earlier steps of the evolution, if they are, and it is 
not evident in advance, whether the last gluon splitting dominates or not.
The goal of out study here is to clarify this point in more detail than in 
~\cite{BLZ}.

\section{Theoretical framework}
\indent

Here we use the specific property of the KMR scheme~\cite{KMRW}
 which enables us to
discriminate between the various components of the quark densities.
We start from the leading order ${\cal O}(\alpha)$ subprocess
"$q^*+\bar{q^*}\to \gamma$", and then divide it into several contributions
which correspond to the interactions of valence quarks
$q_v(x,{\mathbf k}_T^2,\mu^2)$,
sea quarks appearing at the last step of the gluon evolution
$q_g(x,{\mathbf k}_T^2,\mu^2)$,
and sea quarks coming from the earlier steps
$q_s(x,{\mathbf k}_T^2,\mu^2)$.

The KMR approach represents an approximate treatment of the parton
evolution mainly based on the DGLAP equation and incorporating the
BFKL effects at the last step of the parton ladder only, in the form
of the properly defined Sudakov formfactors $T_q(k_t^2,\mu^2)$ and
$T_g(k_t^2,\mu^2)$.
These formfactors already include logarithmic loop correction.
In this approximation, the unintegrated quark distributions
are given by

\begin{figure}[!hb]
  \includegraphics[height=.27\textheight]{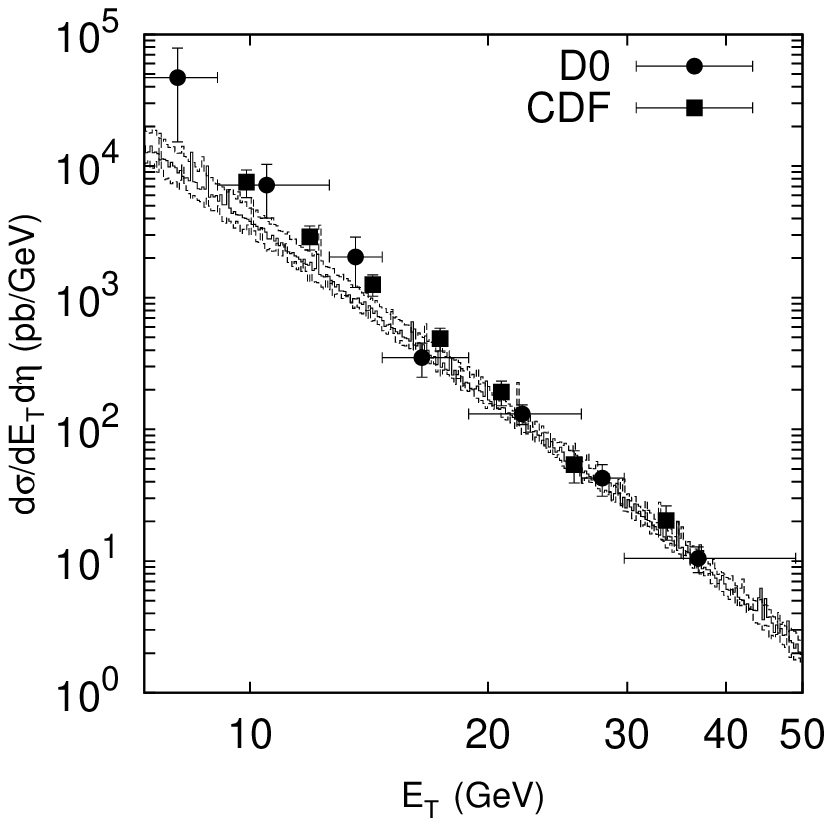}
  \includegraphics[height=.27\textheight]{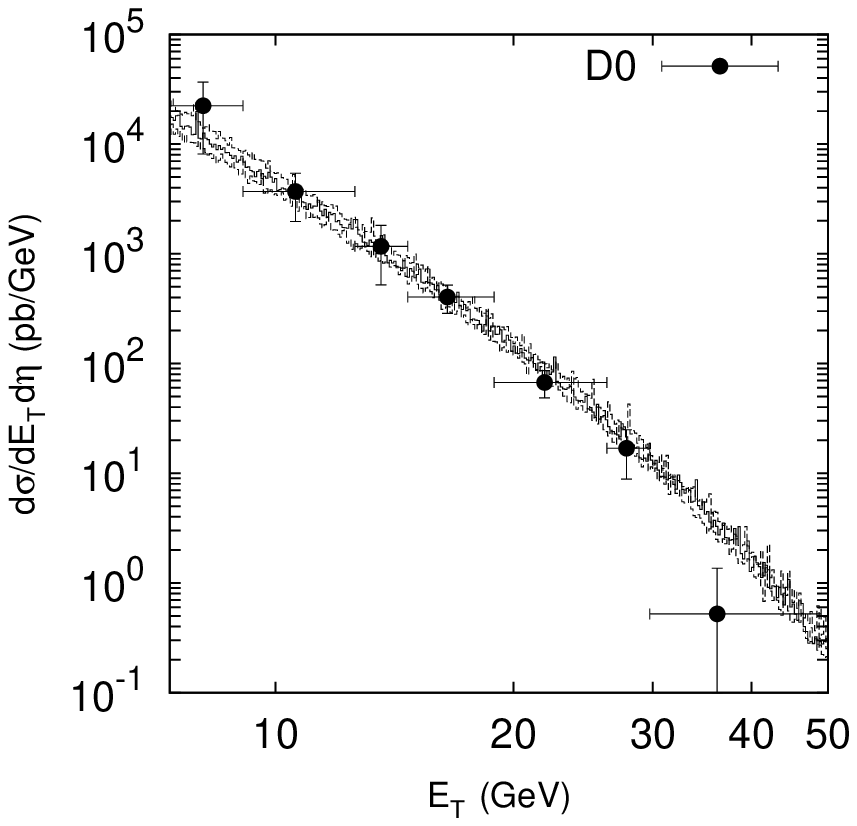}
  \caption{ $d\sigma/dE_Td\eta$ of the inclusive prompt
photon production at $\sqrt s = 630$ GeV
 in the region $|\eta| < 0.9$ 
(left panel) and $1.6 < |\eta| < 2.5$ (right panel). Solid histogram 
corresponds to the hard scale $\mu = E_T$, the upper and the lower 
histograms correspond to the usual variation of the hard scale 
$\mu$.
   }
  \label{fig2}
\end{figure}

\begin{equation}
  \displaystyle f_q(x,{\mathbf k}_T^2,\mu^2) = T_q({\mathbf k}_T^2,\mu^2)
 {\alpha_s({\mathbf k}_T^2)\over 2\pi} \times \atop {
  \displaystyle \times \int\limits_x^1 dz \left[P_{qq}(z)
 {x\over z} q\left({x\over z},{\mathbf k}_T^2\right)
 \Theta\left(\Delta - z\right) + P_{qg}(z)
 {x\over z} g\left({x\over z},{\mathbf k}_T^2\right) \right],}
\label{KMR_q}
\end{equation}
%
where $P_{ab}(z)$ are the usual unregularised leading order DGLAP
splitting functions, and $q(x,\mu^2)$ and $g(x,\mu^2)$ are the
conventional (collinear) quark and gluon densities.
Modifying Eq. (\ref{KMR_q}) in such a way that only the first term is
kept and the second term omitted, we switch the last gluon splitting
off, thus excluding the $q_g(x,{\mathbf k}_T^2,\mu^2)$ component.
 Additional conditions which preserve  our calculations  
from divergences and, also, the isolation cuts which  suppress
the contribution of fragmentation component in the photon production
have been discussed in~\cite{BLZ}.
\begin{figure}[!b]
  \includegraphics[height=.27\textheight]{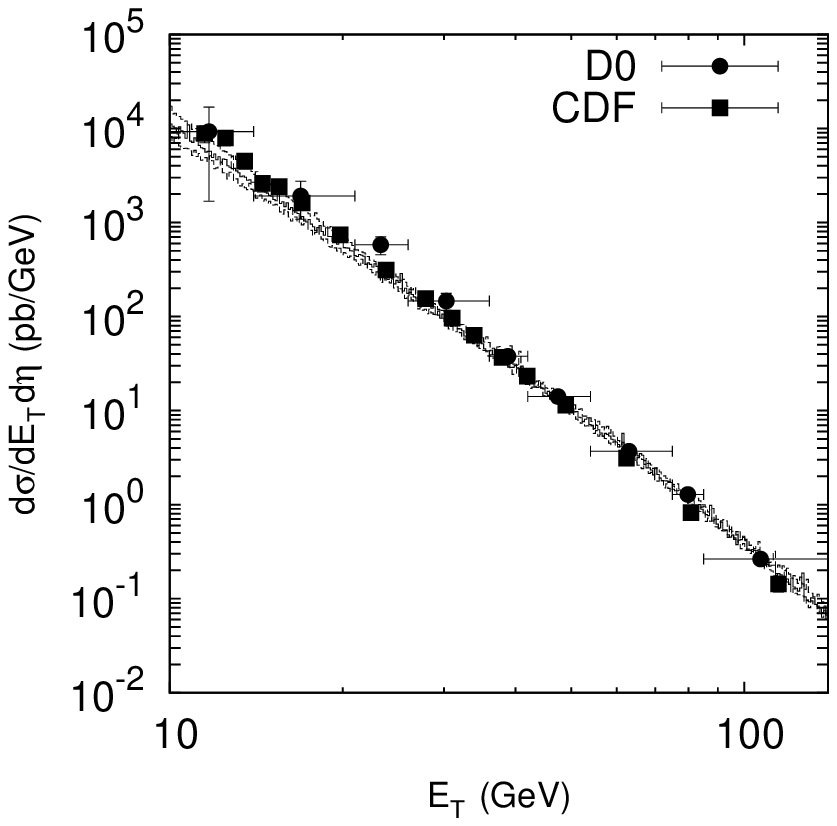}
  \includegraphics[height=.27\textheight]{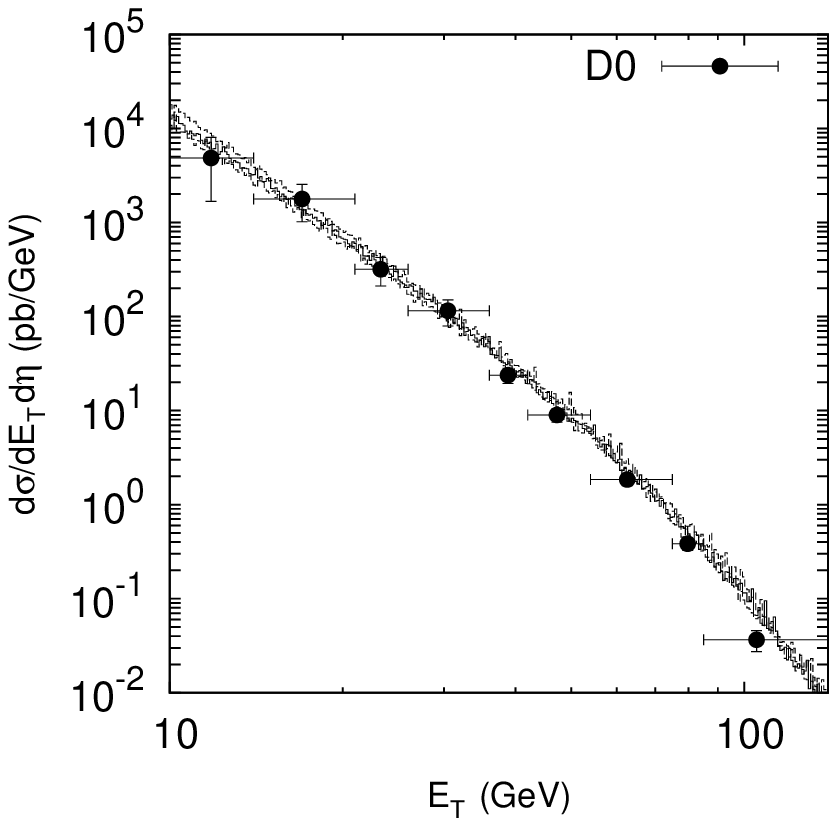}
  \caption{ $d\sigma/dE_Td\eta$ of the inclusive prompt
photon production at $\sqrt s = 1800$ GeV in the region $|\eta| < 0.9$
(left panel) and $1.6 < |\eta| < 2.5$ (right panel). The notation of
the histograms is as in Fig.~\ref{fig2}.
  }
  \label{fig3}
\end{figure}

\section{Numerical results}
\indent

In Fig.~\ref{fig1} we show a comparison between the different 
contributions to the 
inclusive cross section of the prompt photon production at  Tevatron and 
LHC energies. The solid, dashed and dotted histograms represent 
the contributions from the 
$g^* g^* \to \gamma q \bar q, \; q_v  g^* \to
\gamma  q$ and $q_v  \bar q_v \to \gamma   g$ subprocesses, 
respectively. The dash-dotted histograms represent the sum of the 
contributions from the $q_s  g \to \gamma q,\; q_s  \bar q_s \to \gamma
 g$ and $q_v  \bar q_s \to \gamma  g$ subprocesses. Below we denote
it as "reduced sea" component~\footnote{This component has not been
taken into account in~\cite{BLZ}.}. The thick solid histograms represent 
the sum of all contributions. We see that the gluon-gluon fusion 
 is an important production mechanism of the prompt photon production
 both at the Tevatron and LHC conditions. At the LHC, it gives the 
  main contribution to the cross section.
approximately 30\% contribution to the total cross section of prompt 
photon production at Tevatron and approximately 20\% contribution at 
LHC. 
 
Figs.~\ref{fig2} and~\ref{fig3} confront the double differential cross 
sections $d\sigma/
dE_Td\eta$ of the prompt photon production calculated at $\sqrt s = 630$
and 1800 GeV with D$\oslash$~\cite{d0} and CDF~\cite{cdf} data.
 One can see that our results agree very well
with the Tevatron data.
 
In summary we have studied the production of prompt photon in hadronic 
collisions at high energies in the $k_T-$factorization approach of QCD.
The central part of our consideration is off-shell gluon-gluon fusion
subprocess $g^*  g^* \to \gamma  q \bar q$. The contribution from 
the valence quarks has been taken into account additionally.
We demonstrate in the KMR approximation that important contribution to 
total cross sections of
process under consideration also comes from the sea quark interactions.
\section{Acknowledgments}
We thank to DESY Directorate for support (Moscow --- DESY project on MC 
implementation for HERA-LHC), FASI of RF (NS-1456.2008.2) and  RFBR 
(08-02-00896-a). A.L. was supported by the grants of President of 
RF (MK-432.2008.2) and HRJRG.

\bibliographystyle{aipproc}   

\end{document}